\documentclass[epsfig,psfig,aps,twocolumn,prb,showpacs]{revtex4-1}
\usepackage[colorlinks=true,citecolor=red,urlcolor=blue]{hyperref}
\usepackage{amsfonts}
\usepackage{graphicx}
\usepackage{epsfig}
\usepackage{epsf}

\begin{document}
\draft
\title
{
Interlayer magnetoresistance in multilayer Dirac electron systems: motion and merging of Dirac cones
}
\author
{
M. Assili and S. Haddad 
} 
\address{
Laboratoire de Physique de la Mati\`ere Condens\'ee, D\'epartement de Physique,
Facult\'e des Sciences de Tunis, Universit\'e Tunis El Manar, Campus universitaire 1060 Tunis, Tunisia
}
%
%
\begin{abstract}
We theoretically study the effect of the motion and the merging of Dirac cone on the interlayer magnetoresistance in multilayer graphene like systems. This merging, which could be induced by a uniaxial strain, gives rise in monolayer Dirac electron system to a topological transition from a semi-metallic phase to an insulating phase where Dirac points disappear.
Based on a universal Hamiltonian proposed to describe the motion and the merging of Dirac points in two dimensional Dirac electron crystals, we calculate the interlayer conductivity of a stack of deformed graphene like layers using Kubo formula in the quantum limit where only the contribution of the $n=0$ Landau level is relevant.
A crossover from a negative to a positive interlayer magnetoresistance is found to take place as the merging is approached. This sign change of the magnetoresistance could also result from a coupling between the Dirac valleys which is enhanced as the magnetic field amplitude increases. Our results may describe the behavior of the magnetotransport in the organic conductor $\alpha$-(BEDT)$_2$I$_3$ at high pressure where the merging of Dirac cones could be observed.
\end{abstract}
\maketitle

\section{Introduction}
Since the discovery of graphene \cite{Geim,Review}, systems showing massless Dirac electron like dispersion relation continue to attract considerable interest. The signature of such electrons has been recently brought out in the organic conductor \cite{Kobayashi} $\alpha$-(BEDT)$_2$I$_3$ where BEDT stand for bis(ethylenedithio)-tetrathiafulvalene. This compound consists of a stack of conducting BEDT layers separated by the insulating iodine planes. The weak coupling between the conducting layers gives rise to the 2D character for the electronic properties of this material.
Theoretical studies \cite{Kobayashi,Mark-BEDT} and band energy calculations \cite{Kino,Pouget} have given evidence for the presence of tow tilted Dirac cones which move under pressure. 
It has been argued that the interlayer magnetoresistance is a powerfull tool to probe the properties of the Dirac cones \cite{Morinari}.
Tajima {\it et al.} \cite{Tajima} have observed a large negative interlayer magnetoresistance in $\alpha$-(BEDT)$_2$I$_3$ for a transverse magnetic field. This effect was ascribed to the carriers of the zero mode Landau level ($n=0$). 
The authors have also reported a change to a positive magnetoresistance which was assigned to Zeeman splitting of the $n=0$ Landau level.

A theoretical interpretation of these experimental results has been proposed by Osada \cite{Osada08,Osada11} by calculating, within a quantum approach, the interlayer magnetoresistance in a system of stacked Dirac electron layers.
Osada showed that in the quantum limit, the negative magnetoresistance is due to the degeneracy of the zero mode Landau level which dominates the interlayer transport.
At high field, spin splitting becomes relevant and gives rise to the crossover from negative to a positive magnetoresistance due to the reduction of carriers density.
Osada has also given an explanation of the angle dependence of the interlayer resistance observed by Tajima {\it et al.} \cite{Tajima} and which could not be understood within a semi-classical description.\

Based on transport measurements, Monteverde {\it et al.} \cite{Pasquier} have, recently, argued that the conduction in $\alpha$-(BEDT)$_2$I$_3$ could not only be ascribed to Dirac electrons. Both massive and Dirac carriers contribute to the conduction properties. Moreover, no merging of Dirac cones has been observed up to 3 GPa\cite{Pasquier} .\

According to theoretical calculations \cite{Kobayashi11}, the merging in $\alpha$-(BEDT)$_2$I$_3$ is expected to occur around 0.5 GPa. \
More recently, Pi\'echon {\it et al.} \cite{Piechon} discussed, based on an analytical approach, the stability of Dirac points and the merging conditions in $\alpha$-(BEDT)$_2$I$_3$.\

The motion and the merging of Dirac points has been observed in a tunable honeycomb optical lattice of ultracold Fermi gas \cite{Tarruell}. Lim {\it et al.} \cite{Lim} provided a theoretical description of the experiment of Dirac point manipulation in optical lattices.\

The topological phase transition from a zero gap band state, with two Dirac points, to a gaped phase was also observed in a microwave experiment simulating a strained graphene \cite{Bellec}.

Montambaux {\it et al.}\cite{Gilles09} have proposed a universal Hamiltonian to describe the motion and the merging of Dirac cones in 2D systems. The proposed Hamiltonian offers the possibility the follow continuously the toplogical transition from the semi-metallic phase, with two Dirac points, to the insulating phase where the Dirac points merge and a gap opens in the energy spectrum.\

It is worth to stress that the merging in a monolayer graphene could not practically be observed since a very large strain is required \cite{Gilles09,Mark}. It is then interesting to investigate the possible signature of this merging in other Dirac electron systems, in particular in multilayer graphene like system such as the organic conductor $\alpha$-(BEDT)$_2$I$_3$.\\ 

In this paper, we propose to study the merging of Dirac cones in a stack of undoped Dirac electron layers weakly coupled by a vertical tunneling. 
We look for the evidence of the merging in the behavior of the interlayer magnetoresistance.
In the present work, we consider that each layer is described by the universal Hamiltonian proposed by Montambaux {\it et al.}\cite{Gilles09} and we introduce the interlayer coupling perturbatively. 
For simplicity, we do not consider the tilt of Dirac cones which has been addressed in Refs.\onlinecite{Morinari,Himura}. Moreover, we neglect the Zeeman effect which is, already, found to induce a sign change in the magnetoresistance \cite{Tajima,Osada08}. We also do not take into account the broadening of the Landau level which may result in a mixing of the Landau levels \cite{Osada08}.
We derive, based on Kubo formula, the interlayer magnetoresistance in the quantum limit where only the zero mode ($n=0$) is considered. 
In the next section, we focus on the behavior of the field and angle dependence of the interlayer magnetoresistance far from the merging of Dirac cones.
In section III, we derive the magnetoresistance at the merging and discuss its experimental fingerprints.

\section{Interlayer magnetoresistance: motion of Dirac cones}
\subsection{Independent Dirac valleys}
\begin{figure}[hpbt]  
\begin{center}
\vspace{0.5cm}
\includegraphics[width=0.4\columnwidth]{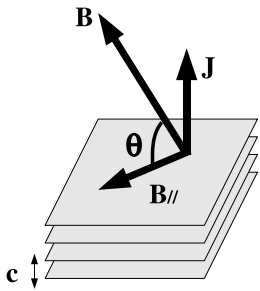}
\end{center}
\caption{Schematic representation of a multilayer system with magnetic field and current configuration. $c$ denotes the interlayer distance and $\theta$ is the out of plane angle of the magnetic field from the conducting layer.} 
\label{Figlayer}
\end{figure}

We consider a stacking structure of layers weakly coupled along the transverse direction and we denote by $t_c$ the interlayer tunneling parameter (Fig. \ref{Figlayer}).
Each layer is a grahene like system described by a triangular lattice with two atoms A and B by a unit cell.\\

%

\begin{figure}[hpbt] 
\begin{center}
\vspace{0.5cm}
\includegraphics[width=0.7\columnwidth]{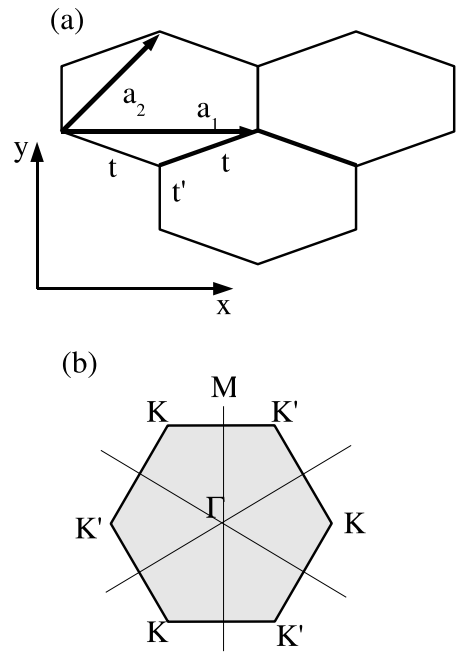}
\end{center}
\caption{(a) Deformation of the honeycomb lattice along the $y$ direction. ($\vec{a}_1,\vec{a}_2$) is the lattice basis. The hopping parameters to the first neighbors $t$ and $t^{\prime}$ are different due the deformation. (b) Brillouin zone of undeformed graphene lattice. By applying a deformation, Dirac cones leave the $K$ and $K^{\prime}$ points and move in the same direction and eventually merge in $M$ point \cite{Mark}.} 
\label{FigZB}
\end{figure}

In the absence of lattice deformation, the Dirac points $\vec{D}$ and $\vec{D}^{\prime}=-\vec{D}$ are in the points $K$ and $K^{\prime}$ at the corners of the first Brillouin zone (BZ) \cite{Gilles09} (Fig. \ref{FigZB}).
By applying a uniaxial strain, along the $y$ direction for example, the Dirac points leave the corners of the BZ and move into the same direction to merge in a single point $\vec{D}_0$ denoted $M$ in Fig.\ref{FigZB}.\\

To describe the motion and the merging of Dirac cones in zero magnetic field, Montambaux {\it et al.}\cite{Gilles09} have proposed the following Hamiltonian, so called, universal Hamiltonian
\begin{eqnarray}
H_0(\vec{p})= 
\left(
\begin{array}{cc}
    0 & \Delta+ {p_x^2 \over 2 m^{\ast}} -  i c_y p_y  \\
 \Delta + {p_x^2 \over 2 m^{\ast}} +  i c_y p_y & 0 \\
  \end{array}
\right) 
\label{UH0} 
\end{eqnarray}
where $\vec{p}=(p_x,p_y)$ is the momentum measured relatively to the merging point $\vec{D}_0$, $c_y$ is the electron velocity along the $y$ direction, $m^{\ast}$ is an effective mass supposed to be positive and $\Delta$ is the parameter governing the topological transition.
This two band Hamiltonian is written in the basis of the A and B site atom eigenstates $(\psi_A,\psi_B)$. \\

The universal Hamiltonian of Eq.\ref{UH0} describes a deformed graphene sheet in the presence of a uniaxial strain applied along the $y$ axis \cite{Mark}.
The corresponding energy spectrum is:
\begin{equation}
\epsilon=\pm \sqrt{\left(\Delta + {p_x^2 \over 2 m^{\ast}}\right)^2 + p_y^2 c_y^2}
\label{eps} 
\end{equation}.
Equation \ref{eps} shows an hybrid character, called by Montambaux {\it et al.}\cite{Gilles09} a semi-Dirac spectrum, with a Schr\"{o}dinger like behavior along the $x$ direction and a Dirac structure along the $y$ axis.\\

The case of $\Delta<0$ corresponds to two distinct Dirac points along the $x$ axis at $\pm p_D$ where $p_D=\sqrt{-2m^{\ast}\Delta}$ whereas for $\Delta=0$, the Dirac points merge at $\vec{D}_0$.\

For $\Delta>0$, a gap of $2\Delta$ opens in the energy spectrum and the system becomes insulator.\newline
To recover the full Dirac spectrum of undeformed graphene, the spectrum given by Eq.\ref{eps} can be linearized along the $x$ axis in the vicinity of Dirac points \cite{Gilles09}.\\.

Let us now focus on the case where the deformed system is in the presence of a magnetic field $$\vec{B}(B_x=B\cos \theta \cos \phi,B_y=B\cos \theta \sin \phi,B_z=B\sin \theta)$$. The Hamiltonian given by Eq.\ref{UH0} can be written, using the Peierls substitution $\vec{p}\rightarrow \vec{p}+e\vec{A}$, as:
\begin{eqnarray}
H(\vec{\pi})= 
\left(
\begin{array}{cc}
    0 & \Delta+ {\pi_x^2 \over 2 m^{\ast}} -  i c_y \pi_y  \\
 \Delta + {\pi_x^2 \over 2 m^{\ast}} +  i c_y \pi_y & 0 \\
  \end{array}
\right) 
\label{Hpi} 
\end{eqnarray}

where the effective momentum is given by $\vec{\pi}=(\pi_x=p_x+ezB_y-eyB_z,\pi_y=p_y-ezB_x,0)$ within the gauge $\vec{A}=(zB_y-yB_z,-zB_x,0)$. $\pi_x$ and $\pi_y$ satisfy the commutation relation $\left[\pi_x,\pi_y\right]=-ie\hbar B_z$.\

To derive the eigenfunctions and the energy spectrum of the Hamiltonian given by Eq.\ref{Hpi}, we consider for simplicity, as in Ref.\onlinecite{Gilles09}, the squared Hamiltonian $H_{eff}$. The eigenproblem reduces to:
\begin{eqnarray}
H_{eff}
\left(
\begin{array}{c}
   \psi_A \\
 \psi_B 
  \end{array}
\right)
=E^2_n
\left(
\begin{array}{c}
    \psi_A \\
 \psi_B 
 \end{array}
\right)
\end{eqnarray}
which may be written as:
\begin{widetext}
\begin{eqnarray}
\left\lbrace \left(\Delta+\frac{\pi^2_x}{2m^{\ast}}\right)^2+c^2_y\pi^2_y+is\frac{c_y}{2m^{\ast}}\left[\pi^2_x,\pi_y\right]\right\rbrace \psi_{A,B}= E^2_n\psi_{A,B}
\label{eigen1}
\end{eqnarray}
\end{widetext}
where $s=\pm$ corresponds respectively to the A and the B sites.
Equation \ref{eigen1} takes the form:
\begin{eqnarray}
\left[  c^2_y\left(P^2_y+e^2z^2B^2_x-2ezB_xP_y\right)+V(Y)\right]\psi_{A,B}= E^2_n\psi_{A,B}
\label{eigen2}
\end{eqnarray}
The potential $V(Y)$ is given by 
\begin{equation}
V(Y)=\left(\frac{e^2B^2_z}{2m^{\ast}}\right)^2\left(\delta+Y^2\right)^2+s\frac{c_y\hbar e^2B^2_z}{m^{\ast}}Y 
\label{V(y)}
\end{equation}
Here $Y=sy_0-y$, $y_0=\frac{p_x+ezB_y}{eB_z}$ and we introduce, as in Ref.\cite{Gilles09}, the parameter $\delta=\frac{2m^{\ast}}{e^2B^2_z}\Delta$.\

$V(Y)$ has two minima at $Y_0=\pm\sqrt{|\delta|}$ separated by a distance of $2\sqrt{|\delta|}$ which decreases as the transverse component $B_z$ of the magnetic field increases. \\

$V(Y)$ can be described by a two independent well for large $|\delta|$ corresponding to the case where Dirac cones are far from the merging point. 
In this case, an expansion of $V(Y)$ around $Y_0$ yields to
\[
V(Y)\sim V(u)=4\left(\frac{e^2B^2_z}{2m^{\ast}}\right)^2|\delta|u^2+s\frac{\hbar c_ye^2B^2_z}{m^{\ast}}\sqrt{|\delta|}
\]
here $u=Y-Y_0$ and $|u|\ll\sqrt{|\delta|}$.
The eigenvalues of Eq.\ref{eigen2} are then:
\begin{eqnarray}
 E_n=\pm\frac{\sqrt{2}\hbar c_y}l\sqrt{n}
\end{eqnarray}
where $l=\left(\frac{\hbar m^{\ast}c_y}{\sqrt{|\delta|}e^2B^2_z}\right)^
{\frac 12}$. \

$E_n$ can be written in the form \cite{Gilles09} $E_n=\pm\sqrt{2\hbar c_yc_x eB_z{n}}$  as found by Himura {\it et al.} \cite{Himura} in $\alpha$-(BEDT)$_2$I$_3$ in the case of non tilted Dirac cones.\\

In the limit of large negative $\delta$, and using the notation of Ref.\cite{Osada08}, the eigenstates of Eq.\ref{eigen2} corresponding to the layer position $z_i$ take the following form :\

For the zero mode ($n=0$)
\begin{eqnarray}
 F_{0,y_0,z_i}(\vec{r})=
\left(
\begin{array}{c}
 0 \\
h_{0,y_0,z_i}(\vec{r}) 
\end{array}
\right)
\end{eqnarray}
For $n>0$:
\begin{eqnarray}
 F_{\pm n,y_0,z_i}(\vec{r})=
\frac 1{\sqrt{2}}\left(
\begin{array}{c}
    \pm h_{n-1,y_0,z_i}(\vec{r}) \\
 h_{n,y_0,z_i}(\vec{r}) 
 \end{array}
\right)
\end{eqnarray}
where 
\begin{widetext}
\begin{eqnarray}
h_{n,y_0,z_i}(\vec{r}) = 
\frac 1{\sqrt{L}}\mathrm{exp}\left(i\frac{ezB_x}{\hbar}y\right) \,
\mathrm{exp} \left[ i\frac{eB_z}{\hbar}\left(z_i\frac{B_y}{B_z}\mp\sqrt{|\delta|}
-\tilde{y}_0\right)x\right]u_n(y-\tilde{y}_0)\delta_{z,z_i}
\end{eqnarray}
\end{widetext}
Here $L$ is the layer length along the $x$ direction, $\tilde{y}_0$ is the center coordinate of the harmonic oscillator and $u_n(y-\tilde{y}_0)$ is the corresponding eigenfunction.\\

We now introduce the interlayer hopping Hamiltonian $\Delta H=-2t_c \cos \frac{cp_z}{\hbar}$ as a perturbation \cite{Osada08}. $c$ denotes the interlayer distance.\

From Kubo formula, and to the lowest order in $t_c$, the interlayer conductivity $\sigma_{zz}$ is given by:
\begin{widetext}
 \begin{eqnarray}
\sigma_{zz}(\omega)=\frac {i\hbar}{ V} \sum_{spin}\sum_{n^{\prime},\tilde{y}^{\prime}_0,z^{\prime}_i}\sum_{n,\tilde{y}_0,z_i}
-\frac{f(E_{n^{\prime}})-f(E_n)}{E_{n^{\prime}}-E_n}
\frac{|\langle n^{\prime}, \tilde{y}^{\prime}_0,z^{\prime}_i| \hat{J}_z|n,\tilde{y}_0,z_i\rangle |^2}{\hbar \omega +i{\hbar}{\tau} + E_n-E_{n^{\prime}}}
\label{Kubo}
 \end{eqnarray}
\end{widetext}
where $\hat{J}_z$ is the interlayer current density $\hat{J}_z=\frac {ie}{\hbar}\left[z,\Delta H\right]$ and $\tau$ is the relaxation time.\\

In the quantum limit, the mixing of Landau levels could be neglected and the relevant contribution to the interlayer conduction is due to the zero mode Landau level $n=0$.\
The matrix element of the $\hat{J}_z$ can be expressed in term of the effective interlayer hopping parameter $\tilde{t}_c$ \cite{Osada08}:
\begin{widetext}
\begin{eqnarray}
\langle 0, \tilde{y}^{\prime}_0,z^{\prime}_i| \hat{J}_z|0,\tilde{y}_0,z_i\rangle
=\frac {-ie}{\hbar}  \tilde{t}_c \left[\delta_{z^{\prime}_i,z_i+c}
\delta_{\tilde{y}^{\prime}_0,\tilde{y}_0+c\frac{B_y}{B_z}}
+\delta_{z^{\prime}_i,z_i-c}
\delta_{\tilde{y}^{\prime}_0,\tilde{y}_0-c\frac{B_y}{B_z}}\right]
\end{eqnarray}
\end{widetext}
where
\begin{widetext} 
\begin{eqnarray}
 \tilde{t}_c =t_c\mathrm{exp}\left[i\frac{eB_x}{\hbar}(z^{\prime}_i-z_i)
\left(\frac{\tilde{y}^{\prime}_0+\tilde{y}_0}2\right)\right]
\mathrm{exp}\left[-\frac{ec^2}{4\hbar B_z}\left(\frac{c_y}{c_x}B^2_x+\frac{c_x}{c_y}B^2_y\right)\right]
\label{tceff}
\end{eqnarray}
\end{widetext}
The interlayer DC conductivity $\sigma_{zz}=\Re\left( \sigma_{zz}(\omega=0)\right) $ can then be written as \cite{Osada08}:
\begin{eqnarray}
\sigma_{zz}=2C\frac {t^2_c e^3c\tau|B_z|}{ \pi\hbar^3}
\mathrm{exp}\left[-\frac 12 \frac{ec^2}{\hbar B_z}\left(\frac 1{\alpha^2}B^2_x
+\alpha^2B^2_y\right)\right]
\label{sigma}
\end{eqnarray}
where the factor $C$ given in Ref.\cite{Osada08} could be considered as a constant as far as the relaxation $\tau$ is assumed to be field independent \cite{Osada08}. \newline
We denote by $\alpha$ the parameter measuring the amplitude of the strain:
$\alpha=\sqrt{\frac{c_x}{c_y}}$, $c_x=\sqrt{\frac{-2\Delta}{m^{\ast}}}$. The same parameter has been introduced by Himura {\it et al.} \cite{Himura} as a measure of the the anisotropy strength. In our calculation, based on the universal Hamiltonian, this parameter describes the proximity of Dirac cones to the merging point. The smaller $\alpha$, the closer the merging.\

In Eq.\ref{sigma}, the proportionality of the prefactor to $B_z$ is due to the Landau level degeneracy $\frac 1{2\pi l^2_B}=\frac{eB_z}{2\pi\hbar}$.\

Equation \ref{sigma} is similar to the interlayer conductivity expression obtained by Osada \cite{Osada08} if one takes the case of undeformed graphene like system $\alpha=1$.\

It is worth to note that the expression of $\sigma_{zz}$ given by Eq.\ref{sigma} is also reminiscent of that obtained by Himura {\it et al.} \cite{Himura} in $\alpha$-(BEDT)$_2$I$_3$ in the case of non tilted Dirac cones. \\

\begin{figure}[hpbt] 
\begin{center}
\vspace{0.5cm}
\includegraphics[width=0.7\columnwidth]{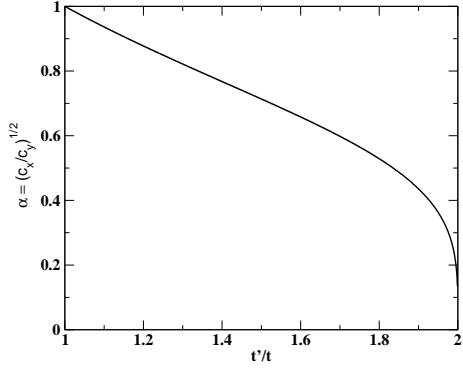}
\end{center}
\caption{Dependence of the anisotropy parameter $\alpha=\sqrt{\frac{c_x}{c_y}}$ as a function of the hopping parameter ratio $\frac{t^{\prime}}t$. At the merging, $\alpha$ vanishes.} 
\label{Figalpha}
\end{figure}

If the hopping integrals are limited to the first neighboring atoms, the universal Hamiltonian parameters, in the case of fixed $\Delta$ and $c_y$, take the form \cite{Gilles09}:
\begin{eqnarray}
 \Delta=t^{\prime}-2t,\, c_x=\frac{\sqrt{3}a_0}{\hbar}\sqrt{t^2-\frac{t^{\prime 2}}4},\,
\mathrm{and}\, c_y=\frac{3t^{\prime}a_0}{2\hbar}
\end{eqnarray}
where $t^{\prime}$ is the hopping parameter along the $y$ direction which is different from the other first neighbor hopping parameter $t$ regarding the effect of the uniaxial strain (Fig.\ref{FigZB}). $a_0$ is the distance between the two atoms of the unit cell.\
The dependence of $\alpha$ on the hopping parameter $t^{\prime}$ is represented in Fig.\ref{Figalpha}. At the merging, $t^{\prime}=2t$, $\alpha$ vanishes.\

To derive the interlayer conductivity given by Eq.\ref{sigma}, we have assumed a large negative $\Delta$ so that the two valleys of the double well potential $V(Y)$ (Eq.\ref{V(y)}) could be considered as independent. The two Dirac cones are far from the merging point $\vec{D}_0$.
This assumption could be justified as far as $\alpha$ is not close to zero. One needs to define a criterion to fix the critical value of $\alpha$ below which the two potential valleys start to interact. This point will be discussed later.\\

According to Eq.\ref{sigma}, $\sigma_{zz}$ increases linearly with the field amplitude for a normal magnetic field as found in the undeformed case \cite{Osada08}.\

\begin{figure}[hpbt] 
\begin{center}
\vspace{0.5cm}
\includegraphics[width=0.7\columnwidth]{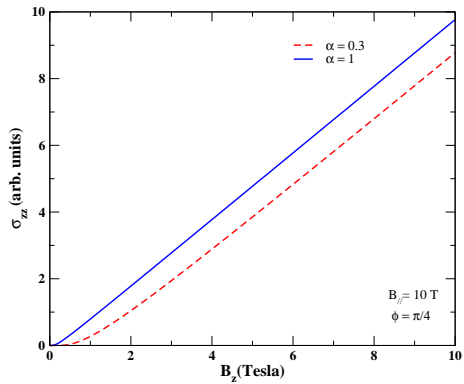}
\end{center}
\caption{Interlayer conductivity as a function of the normal component of the magnetic field at a fixed inplane field component.} 
\label{sigmaBz}
\end{figure}

In the presence of an in-plane field component $B_{\parallel}$, the increase of $\sigma_{zz}$ is reduced compared to the case of undeformed layer as shown in Fig.\ref{sigmaBz}. This decrease is due to the Gaussian decay of the effective interlayer tunneling amplitude $\tilde{t}_c$ (Eq.\ref{tceff}).\
The inplane component  $B_{\parallel}$ generates a positive magnetoresistance effect since it induces an inplane Lorentz force. The latter reduces the interlayer tunneling giving rise to a positive magnetoresistance.\\

\begin{figure}[hpbt] 
\begin{center}
\vspace{0.5cm}
\includegraphics[width=0.7\columnwidth]{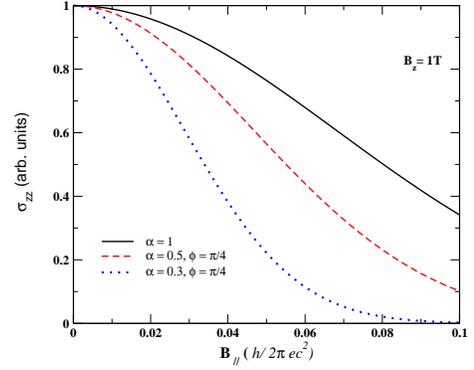}
\end{center}
\caption{Interlayer conductivity as a function of the inplane component of the magnetic field expressed in the unit of $\frac{\hbar}{ec^2}$. } 
\label{sigmaB}
\end{figure}

According to Fig.\ref{sigmaB}, the decrease of magnetoresistance with$B_{\parallel}$ is more pronounced as $\alpha$ decreases.\

\begin{figure}[hpbt] 
\begin{center}
\vspace{0.5cm}
\includegraphics[width=0.7\columnwidth]{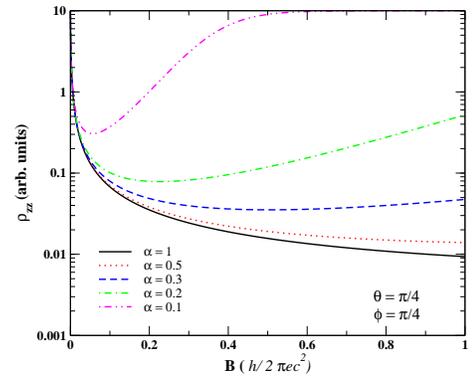}
\end{center}
\caption{Interlayer resistivity $\rho_{zz}$ as a function of the field amplitude for different value of the anisotropy parameter $\alpha=\sqrt{\frac{c_x}{c_y}}$.
Calculations are done for $\theta=\frac{\pi}4$ and $\phi=\frac{\pi}4$ and the filed is expressed in the unit of $\frac{\hbar}{ec^2}$.} 
\label{rho}
\end{figure}

In figure \ref{rho}, we plot the dependence of the interlayer resistivity $\rho_{zz}$ as a function on the magnetic field amplitude $B$. $\rho_{zz}$ is given by \cite{Osada08,Morinari,Himura}
\begin{eqnarray}
\rho_{zz}=\frac A {B_0+|B_z|\mathrm{exp}\left[-\frac 12 \frac{ec^2}{\hbar B_z}\left(\frac 1{\alpha^2}B^2_x
+\alpha^2B^2_y\right)\right]}
\label{rhozz}
\end{eqnarray}
Here $B_0=0.1$ T is a fitting parameter \cite{Osada08} and  $A=\frac{\pi\hbar^3\tau}{2Ct^2_ce^3c}$.\\

Figure \ref{rho} shows a negative magnetoresistance for $\alpha$ close to unity.
It turns out that the negative magnetoresistance feature predicted in multilayer Dirac electron systems \cite{Osada08,Himura} is a robust effect which survives under uniaxial strain.\\

However, as $\alpha$ decreases, the Gaussian exponential decay in $\sigma_{zz}$ (Eq.\ref{sigma}) overcomes the linear increase with $B_z$ and a crossover from negative to positive interlayer magnetoresistance takes place at a critical field value $B_{cr}(\alpha)$. The latter is found to decrease as $\alpha$ is reduced. \
This sign change of the magnetoresistance is due to a strain renormalization of the inplane field component $B_{\parallel}$
According to Eq.\ref{sigma}, $B_{\parallel}$ is responsable of the Gaussian exponential decay giving rise to a positive magnetoresistance.
This effect is enhanced in the presence of the anisotropy which renormalizes the inplane field component: the effective field component, along the direction of Dirac point motion, $B_x$ is renormalized as $\frac{B_x}{\alpha}$ which gets larger as the merging is approached ($\alpha$ decreases).\\

\begin{figure}[hpbt] 
\begin{center}
\vspace{0.5cm}
\includegraphics[width=0.7\columnwidth]{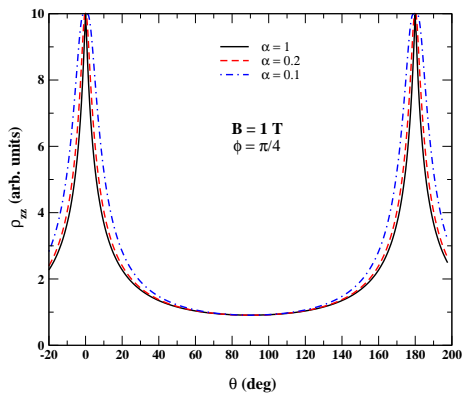}
\end{center}
\caption{Out of plane angle dependence of the interlayer resistivity at a fixed field amplitude for different values of $\alpha$ parameter.} 
\label{Figtheta}
\end{figure}

In figure \ref{Figtheta} we present the dependence of the interlayer resistance on the field orientation from the conducting plane at fixed field amplitude and for different $\alpha$ values.
The maximum of the resistivity is, as in undeformed graphene layers \cite{Osada08}, for an inplane magnetic field ($\theta=0$).
The results show that the peak for an inplane field is broadned as $\alpha$ decreases reflecting the enhancement of the inplane Lorentz force.\
This behavior is found to be unchanged by varying the azimuthal angle $\Phi$.\

\begin{figure}[hpbt] 
\begin{center}
\vspace{0.5cm}
\includegraphics[width=0.7\columnwidth]{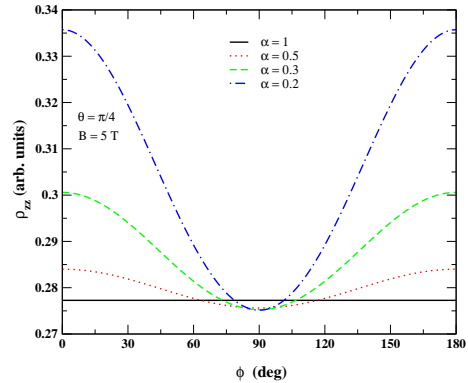}
\end{center}
\caption{Dependence of the interlayer resistivity on the azimuthal angle $\phi$ at a fixed field amplitude for different values of $\alpha$ parameter.} 
\label{Figphi}
\end{figure}

The azimuthal angle dependence of the interlayer resistance for different values of $\alpha$ is depicted in Fig.\ref{Figphi} which shows that the motion of Dirac cones ($\alpha\neq 1$)gives rise to the $\Phi$ dependence of $\rho_{zz}$. The maximum of the resistivity corresponds to $\Phi=0$ where the inplane magnetic field component is aligned perpendicular to the strain direction $y$.\

Morinari {\it et al.} \cite{Morinari} ascribed the $\Phi$ dependence of $\rho_{zz}$ in $\alpha$-(BEDT)$_2$I$_3$ to the tilt of Dirac cones rather than to the anisotropy of Fermi velocity  .
We propose that a $\Phi$ dependent interlayer magnetoresistance could be a signature of Dirac cone motion.
The question is how to distinguish experimentally between the $\Phi$ dependence due to the anisotropy of Fermi velocities (or the tilt of Dirac cones) and that induced by Dirac point motion?\

Fermi surface probes may be a key issue to unveil this question.\\

To summarize, the magnetoresistance shows, in the case of independent Dirac valleys, a $\phi$ dependence for deformed honeycomb lattice and a possible sign change in the presence of an inplane field component. The latter gives rise to an inplane Lorentz force which gets enhanced as the strain amplitude increases  ($\alpha$ decreases). This enhancement results from a strain renormalization of the field component along the direction perpendicular to the deformation axis.\\

At this point a natural question arises: below which value of $\alpha$ our assumption of independent Dirac valleys breaks down?
In the following section, we try to bring some answers.\

\subsection{Interacting Dirac valleys}

Regarding the double well structure of the potential $V(Y)$ (Eq.\ref{V(y)}), the Landau levels are doubly degenerate in the case where $\alpha$ is close to unity (large $\Delta$). However, by approaching the merging (reducing $\alpha$), the degeneracy is removed due to the tunneling between the Dirac valleys. The $E_{n=0}$ level splits into two levels separated by a gap. The system undergoes a crossover from a semi-metallic state, where $E_{n=0}$ is degenerate, to a semi-conducting state where the degeneracy is lifted. This is expected to result in a change form a negative to a positive interlayer magnetoresistance since the carrier density of the lowest energy level is reduced.\\

The valley degeneracy could also be removed, at a fixed value of $\Delta$, by increasing the magnetic field amplitude: the distance between the potential minima (Eq.\ref{V(y)}) decreases with increasing the magnetic field ($2Y_0\propto \frac{\sqrt{|\Delta|}}{B_z}$). A crossover from a negative to a positive magnetoresistance is expected at a critical field $B_{z,cr}(\alpha)$ for a given value of the parameter $\alpha$. \newline
As the merging is approached ($\alpha$ decreases), the two potential valleys are no more independent and a small field value could remove the valley degeneracy. The closer the merging, the smaller $B_{z,cr}$. \\

It is worth to stress that the crossover from the negative to the positive interlayer magnetoresistance observed by Tajima {\it et al.} \cite{Tajima} in $\alpha$-(BEDT)$_2$I$_3$ has been ascribed by Osada \cite{Osada08} to Zeeman energy effect. Morinari and Tohyama \cite{Morinari10} proposed that the sign change of the magnetoresistance in $\alpha$-(BEDT)$_2$I$_3$ is due to Landau level mixing effect which depends on the level broadening.\

In the following we argue that such crossover may be induced by a tunneling between Dirac valleys which takes place at high magnetic field or close to the merging.\\ 

Montambaux {\it et al.} \cite{Gilles09} have estimated the gap $\Delta E_n$ between the two energy levels obtained when the valley degeneracy of a Landau level $E_n$ is lifted. They found that 
\[
 \Delta E_n\sim \mathrm{e}^{-\# \frac{|\Delta|^{3/2}} { {B}_z}}
\]
Since $\alpha^2=\frac {c_x}{c_y}=\sqrt{\frac{2|\Delta|}{m^{\ast}c_y^2}}$,
$\Delta E_n$ takes the form
\begin{eqnarray}
 \Delta E_n\sim \mathrm{e}^{-\sqrt{2} \frac{(m^{\ast}c_y)^{2}} { e B_z}\alpha^6}
\label{gap}
\end{eqnarray}

The expressions of $m^{\ast}$ and $c_y$ for fixed $m^{\ast}$ and $\Delta$ are given by \cite{Gilles09} :
\[
 m^{\ast}=\frac{2\hbar^2}{3ta^2_0}, \quad  c_y=\frac{3ta_0}{\hbar}
\]
where $a_0$ is the distance between neighboring atoms of the honeycomb lattice.\

$\Delta E_n$ could then be written as:
\begin{eqnarray}
 \Delta E_n\sim \mathrm{e}^{-4\sqrt{2} \frac{\alpha^6}{\tilde{B}_z}}
\label{gap2}
\end{eqnarray}

where the dimensionless field is $\tilde{B}_z=\frac{B_zea^2_0}{\hbar}$.

The critical field $B_{z,cr}$ at which the gap $\Delta E_n$ opens should scales, according to Eq.\ref{gap2}, as:
\begin{eqnarray}
 \ln \tilde{B}_{z,cr} \sim 6\ln \alpha 
\label{eqBcr}
\end{eqnarray}

It is interesting to note that the energy gap around the $n=0$ Landau level was also calculated by Esaki {\it et al.} \cite{Esaki} who also found the same exponential behavior as Montambaux {\it et al.}\cite{Gilles09}.\\

To take into account the interaction between the Dirac valleys inducing the degeneracy lifting of the Landau level, we adopt a perturbative approach. We consider
the first order correction to the Landau energy. The corresponding wave function are those given in the previous section which are the zeroth order correction in terms of the perturbation\newline
The energy difference in the Kubo formula of the conductivity in the quantum limit ($n=n^{\prime}=0$) (Eq.\ref{Kubo}) is then replaced by $\Delta E_n$:
\begin{widetext}
\begin{eqnarray}
\sigma_{zz}(\omega)\sim \frac {i\hbar}{ V} \sum_{spin}\sum_{\tilde{y}^{\prime}_0,z^{\prime}_i}\sum_{\tilde{y}_0,z_i}
-\frac{df}{dE}
\frac{|\langle 0, \tilde{y}^{\prime}_0,z^{\prime}_i| \hat{J}_z|0,\tilde{y}_0,z_i\rangle |^2}{\hbar \omega +i{\hbar}{\tau} + \Delta E_n}
\label{Kubo2}
\end{eqnarray}
\end{widetext}
The DC conductivity takes then the form:
\begin{eqnarray}
\sigma_{zz}=\frac {t|B_z|}{A} \frac 1{1+\frac{\tau^2\Delta E^2_n}{\hbar^2}}
\mathrm{exp}\left[-\frac 12 \frac{ec^2}{\hbar B_z}\left(\frac 1{\alpha^2}B^2_x
+\alpha^2B^2_y\right)\right]
\label{sigma-correction}
\end{eqnarray}
where  $A=\frac{\pi\hbar^3\tau}{2Ct^2_ce^3c}$ as given in the previous section. \

To have a rough estimation of the effect of the energy gap $\Delta E_n$ on the conductivity, we make the following approximation for the correction term 
\[
\frac 1{1+\frac{\tau^2\Delta E^2_n}{\hbar^2}}\sim 1+ 
\mathrm{e}^{-4\sqrt{2} \frac{\alpha^6}{\tilde{B}_z}}
\]
where we replaced the energy gap $\Delta E_n$ by its expression given by Eq.\ref{gap2}.\\

\begin{figure}[hpbt] 
\begin{center}
\vspace{0.5cm}
\includegraphics[width=0.7\columnwidth]{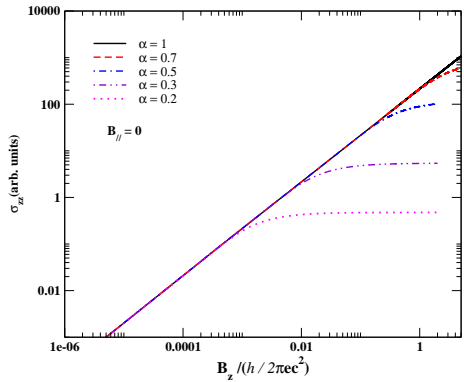}
\end{center}
\caption{Dependence of the interlayer conductivity $\sigma_{zz}$ on the transverse magnetic field $B_z$ for different values of the anisotropy parameter $\alpha$.} 
\label{sigma-coupl}
\end{figure}

In figure \ref{sigma-coupl} we plot the field dependence of the interlayer conductivity $\sigma_{zz}$ (Eq.\ref{sigma-correction}) in the case of a transverse magnetic field. The results show a change in the behavior of $\sigma_{zz}$ at a critical field $B_{z,cr}$ which decreases with decreasing $\alpha$. 

\begin{figure}[hpbt] 
\begin{center}
\vspace{0.5cm}
\includegraphics[width=0.7\columnwidth]{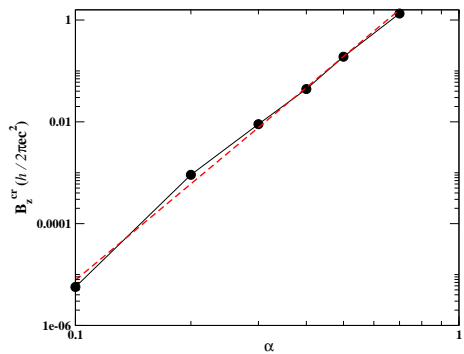}
\end{center}
\caption{$\alpha$ dependence of the critical field $B_{cr}$ at which a change in the behavior of interlayer conductivity takes place (Fig.\ref{sigma-coupl}). The dashed line is a fit of the numerical data showing that $\ln \tilde{B}_{z,cr} \sim 6.3 \ln \alpha$. } 
\label{FigBcr}
\end{figure}

The dependence of $B_{z,cr}$ on $\alpha$ is shown in Fig.\ref{FigBcr} according to which  $\ln \tilde{B}_{z,cr} \sim 6.3 \ln \alpha$ which is in good agreement with the value estimated by Montambaux {\it et al.} \cite{Gilles09} (Eq.\ref{eqBcr}). \
This agreement support the approximation of the first order energy correction we introduced in the Kubo formula (Eq.\ref{Kubo2}) to account for the contribution of the Dirac valley tunneling to the magnetotransport.

It turns out that for a given $\alpha$, our assumption of independent Dirac cones holds as far as $\tilde{B_z}< \alpha^6$ where  $\tilde{B}_z=\frac{B_z}{Bz0}$ with $B_{z0}=\frac{\hbar}{ea^2_0}$. \

In $\alpha$-(BEDT)$_2$I$_3$ compound, $ B_{z0}\sim 660$T since $a_0\sim 10 \AA$ which means that the assumption of independent Dirac cones is justified for a transverse field $B_z< 10 $T for $\alpha=0.5$.\\  

\begin{figure}[hpbt] 
\begin{center}
\vspace{0.5cm}
\includegraphics[width=0.7\columnwidth]{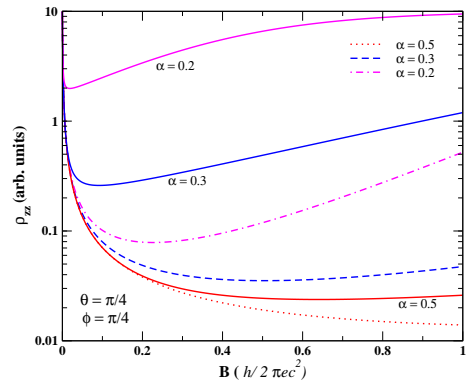}
\end{center}
\caption{Interlayer resistance $\rho_{zz}$ as a function of the field amplitude $B$ for different value of the parameter $\alpha$. The solid (broken) lines correspond to the results with interacting (independent) Dirac valleys. } 
\label{Figrhocoupl}
\end{figure}

In Fig.\ref{Figrhocoupl} we plot the dependence of the magnetoresistance $\rho_{zz}$ on the field amplitude $B$ in the presence of an inplane field component. The solid lines correspond to the results obtained taking into account the coupling between the Dirac valleys whereas the broken lines are the results corresponding to the first section with independent Dirac cones.
Fig.\ref{Figrhocoupl} shows that, in the presence of a coupling between Dirac valleys, the crossover from negative to positive magnetoresistance takes place at a smaller field compared to the case of independent Dirac cones. It turns out that the sign change of the magnetoresistance could be induced by Dirac valleys tunneling  which is enhanced as the magnetic field increases (Eq.\ref{gap2}) or as the merging is approached ($\alpha$ is reduced).\\
  
The outcomes of this section is that the crossover from negative to positive magnetoresistance could be induced by a coupling between Dirac valleys which gets more pronounced by approaching the merging or by increasing the magnetic field amplitude.\

In the following, we focus on the behavior of the interlayer magnetoresistance at the merging and see how the field and the angle dependences of $\rho_{zz}$ are affected.\

\section{Interlayer magnetoresistance: merging of Dirac cones}

The diagonalization of the universal Hamiltonian for $\Delta=0$ and in the gauge $\vec{A}=(zB_y-yB_z,-zB_x,0)$ reduces to
\begin{eqnarray}
 \left(\frac{\hbar c_y}{\gamma}\right)^2\left[  \tilde{\Pi}^2_y+
\tilde{Y}^4-2s\tilde{Y}\right]\psi_{A,B}
 = E^2_n\psi_{A,B}
\label{eigen3}
\end{eqnarray}
where we take, for simplicity, the squared Hamiltonian.
The dimensionless operators $\tilde{\Pi}_y$ and $\tilde{Y}$ are given by: $\tilde{\Pi}_y=\frac{\gamma}{\hbar}\pi_y$, $\tilde{Y}=\frac Y {\gamma}$, $Y=y_0-y$, $y_0=\frac{p_x+ezB_y}{eB_z}$ and $\pi_y=py-ezB_x$.
$\gamma$ is written as \cite{Gilles09} 
\begin{eqnarray}
\gamma=\left(\frac{2\hbar c_ym^{\ast}}{e^2B^2_z}\right)^{\frac 13}
\label{gamma} 
\end{eqnarray}
$\tilde{\Pi}_y$ and $\tilde{Y}$ satisfy the commutation relation $\left[\tilde{Y},\tilde{\Pi}_y\right]=-i$.\

The eigenfunction of Eq.\ref{eigen3} is of the form
\[
 \psi (\vec{r})\sim \rm{e}^{ic_1x}\rm{e}^{ic_2y}\phi(\tilde{Y})
\]
where $c_1=\frac{eB_z}{\hbar}\left(\tilde{y}_0-z\frac{B_y}{B_z}\right)$ and $c_2=\frac{eB_x}{\hbar}z$. $\phi(\tilde{Y})$ is the eigenfunction of  anharmonic quartic oscillator and $\tilde{y}_0$ is the corresponding center coordinate.
$\phi(\tilde{Y})$ is the eigenfunction of:
\[
H_{anh}= \left(\frac{\hbar c_y}{\gamma}\right)^2
\left[\tilde{\Pi}^2_y+\tilde{Y}^4-2s\tilde{Y}\right]
\]
We will consider in the next the case of $s=1$ since the $s=\pm1$ correspond to a symmetric problem as a function of $\tilde{Y}$.\

In Refs.\onlinecite{math1,math2}, the authors studied the eigenproblem of the following anharmonic quartic oscillator
\[
 H_{\pm}=-\frac{d^2}{dx^2}+g^2x^4\pm2g|x|
\]
The groundstate eigenfunction of the potential $V_-(x)= g^2x^4-2g|x|$ is of the form $\phi^-_0(x) \sim \rm{e}^{-g\frac{|x|^3}3}$.\\

Therefore, the solution $\phi(\tilde{Y})$ of the anharmonic part of Eq.\ref{eigen3} can be written, for to the lowest Landau level and for $\tilde{Y}>0$, as:
\begin{eqnarray}
 \phi(\tilde{Y})=\phi^-_0(\tilde{Y})\sim \rm{e}^{-\frac{\tilde{Y}^3}3}
\end{eqnarray}
 
The eigenstate of the zero mode level of Eq.\ref{eigen3} takes then the form:
\begin{eqnarray}
 F_{0,\tilde{y}_0,z_i}(\vec{r})=
\left(
\begin{array}{c}
 0\\
f_{0,\tilde{y}_0,z_i}(\vec{r})
\end{array}
\right)
\end{eqnarray}
where
\begin{widetext}
\[
f_{0,\tilde{y}_0,z_i}(\vec{r})\sim \mathrm{exp}\left[i \frac{eB_z}{\hbar}
\left(\tilde{y}_0-z_i\frac{B_y}{B_z}\right)x\right]
\mathrm{exp}\left[i \frac{exz_iB_x}{\hbar}y\right]
\mathrm{exp}\left[-\frac{\left(\tilde{y}_0-y\right)^3}{3\gamma^3}\right] 
\] 
\end{widetext}
The interlayer hopping matrix is given, as in section II, by:
\begin{widetext}
 \begin{eqnarray}
\langle F_{0,\tilde{y}^{\prime}_0,z^{\prime}_i}| \Delta H|
F_{0,\tilde{y}_0,z_i}\rangle=-\tilde{t}_c(\tilde{y}^{\prime}_0,z^{\prime}_i,\tilde{y}_0,z_i)
\left[ \delta_{z^{\prime}_i,z_i+c}
\delta_{\tilde{y}^{\prime}_0,\tilde{y}_0+c\frac{B_y}{B_z}}
+\delta_{z^{\prime}_i,z_i-c}
\delta_{\tilde{y}^{\prime}_0,\tilde{y}_0-c\frac{B_y}{B_z}} \right]  
 \end{eqnarray}
\end{widetext}

The effective interlayer hopping can then be written as:
\begin{widetext}
\begin{eqnarray}
\tilde{t}_c(\tilde{y}^{\prime}_0,z^{\prime}_i,\tilde{y}_0,z_i)\sim
t_c\exp\left( i \frac{eB_x}{\hbar}\frac{(\tilde{y}^{\prime}_0+\tilde{y}_0)}2
(z^{\prime}_i-z_i)\right) 
\int_0^{\infty} dx \exp\left[ i(z^{\prime}_i-z_i) \frac{eB_x}{\hbar}x\right]
\exp\left[-\frac 13 \left(\frac{x+a}{\gamma}\right)^3\right]
\exp\left[-\frac 13 \left(\frac{x-a}{\gamma}\right)^3\right]
\end{eqnarray}
\end{widetext}
with $x=y-\frac{\tilde{y}^{\prime}_0+\tilde{y}_0}2$, $a=\frac{\tilde{y}^{\prime}_0-\tilde{y}_0}2=\frac {z^{\prime}_i-z_i}2\frac{B_y}{B_z}$ and $\gamma$ is given by Eq.\ref{gamma}.\\

Using the Kubo formula, we obtain the interlayer conductivity to the lowest order contribution of $t_c$:
\begin{eqnarray}
\sigma_{zz}\sim t^2_c |B_z|\left|\int _0^{\infty} \exp\left[ ic\frac{eB_x}{\hbar}x\right] f(x)\right|^2
\label{sigma2}
\end{eqnarray}
where $c$ is the interlayer distance and $f(x)$ is given by:
\begin{widetext}
\begin{eqnarray}
 f(x)=\exp\left[ -\frac 13 \left(\frac c {\gamma}\right)^3
\left(\frac x c +\frac {B_y}{2B_z}\right)^3\right] 
\exp\left[ -\frac 13 \left(\frac c {\gamma}\right)^3
\left(\frac x c -\frac {B_y}{2B_z}\right)^3\right]
\end{eqnarray} 
\end{widetext}

The normal field component $B_z$ appearing in the prefactor of $\sigma_{zz}$ in Eq.\ref{sigma2} is due to the degeneracy of the Landau level.\\

In graphene like system, at the merging ($\Delta=0$), $m^{\ast}=\frac{2\hbar^2}{3ta^2_0}$ and $c_y=\frac{3ta_0}{\hbar}$ where $a_0$ is the distance between the two atoms of the unit cell\cite{Gilles09}.
We then obtain $\left(\frac c {\gamma}\right)^3=\frac {a_0}{4c}\tilde{B}^2_z$, where the dimensionless magnetic field is $\tilde{B}_z=\frac{ec^2}{\hbar}B_z$.\
We take for numerical calculations $c=1.75$nm and $a_0=10\AA$ as in $\alpha$(BEDT)$_2$I$_3$.\\

\begin{figure}[hpbt] 
\begin{center}
\vspace{0.5cm}
\includegraphics[width=0.7\columnwidth]{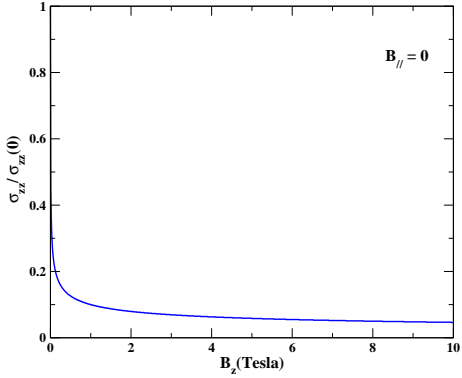}
\end{center}
\caption{Interlayer conductivity as a function of the normal component of the magnetic field at the merging of Dirac points.} 
\label{FigsigmaBz}
\end{figure}

The field dependence of the interlayer conductivity $\sigma_{zz}$ for a transverse magnetic field is plot in Fig.\ref{FigsigmaBz}.

Contrary to the case of separated Dirac cones, $\sigma_{zz}$ decreases, at the merging by increasing the normal field amplitude. This decrease appears as the continuity of the positive interlayer magnetoresistance obtained beyond the crossover field $B_{z,cr}$ for vanishing $\alpha$ in section II.
We can then conclude that the crossover from negative to positive magnetoresistance is due to Dirac cone motion. As $\alpha=\sqrt{\frac{c_x}{c_y}}$ decreases, Dirac cones get closer to the merging point and the interlayer conductivity is reduced.\

We do not claim that our calculations provide a continuous description of the Dirac point motion from the zero gap to the gaped phase. But one could notice that there is a sign change of the magnetoresistance, for a normal field ($B=B_z)$, if the system moves from the independent Dirac valleys to the merging phase.
In the former case, $\sigma_{zz}$ is linear to $B_z$ whereas in the latter case $\sigma_{zz}$ decreases with increasing $B_z$.\

To obtain a continuous analytical description of the magnetotransport with the Dirac point motion, one need to derive the expression of the eigenfunctions of the anharmonic quartic oscillator with the potential given by Eq.\ref{V(y)}.
To the best of our knowledge, these eigenfunctions have not been analytically determined.
Sk\'ala {\it et al.}\cite{Skala} found solutions for the Schr\"odinger equation corresponding to the Hamiltonian
\[
H=-\frac{d^2}{dx^2}+V(x)
\]
where $V(x)=V_1x+V_2x^2+V_3x^3+V_4x^4$ with the condition $V_4>0$.
The solution is of the form:
\[
 \psi(x)=\exp\left( -g_0x-g1x^2/2-g_2x^3\right) 
\]
However, $\psi(x)$ diverges for $x\rightarrow \pm \infty$.\\

\begin{figure}[hpbt] 
\begin{center}
\vspace{0.5cm}
\includegraphics[width=0.7\columnwidth]{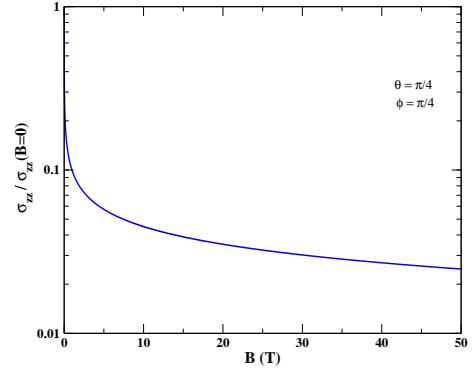}
\end{center}
\caption{Interlayer conductivity as a function of the magnetic field amplitude at the merging of Dirac points.} 
\label{FigsigmaB}
\end{figure}

In Fig.\ref{FigsigmaB} we present the field dependence of the interlayer conductivity $\sigma_{zz}$ in the presence of an inplane field component. $\sigma_{zz}$ shows a positive magnetoresistance as a function of the field amplitude.\\

\begin{figure}[hpbt] 
\begin{center}
\vspace{0.5cm}
\includegraphics[width=0.7\columnwidth]{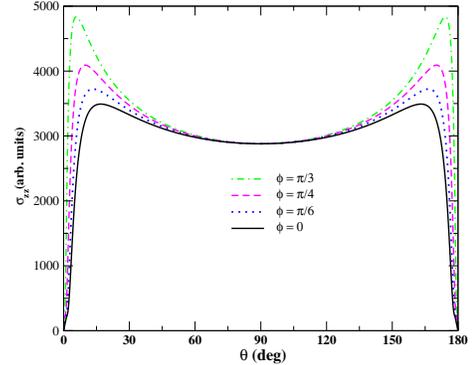}
\end{center}
\caption{Dependence of the interlayer conductivity on the out of plane angle $\theta$ at the merging of Dirac points.} 
\label{Figsigma-teta}
\end{figure}

\begin{figure}[hpbt] 
\begin{center}
\vspace{0.5cm}
\includegraphics[width=0.7\columnwidth]{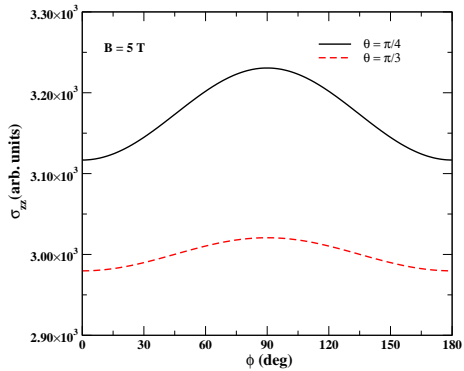}
\end{center}
\caption{Dependence of the interlayer conductivity on the azimuthal angle $\phi$ for a fixed field amplitude at the merging of Dirac points.} 
\label{Figsigma-phi}
\end{figure}

Figures \ref{Figsigma-teta} and \ref{Figsigma-phi} show the field orientation dependence of the interlayer conductivity at the merging.

According to figure \ref{Figsigma-phi} the interlayer conductivity $\sigma_{zz}$ shows a maximum along the strain direction $y$ ($\Phi=\frac{\pi}2$). This behavior is found to be independent of the out-of-plane angle $\theta$ as in the case where Dirac cones are far from the merging (Fig.\ref{Figphi}).\\

However, the dependence of $\sigma_{zz}$ on the out of plane angle $\theta$ is different from that found far from the merging. The maximum of the conductivity is no more for a transverse magnetic field ($\theta=\frac{\pi}2$), as in Fig.\ref{Figtheta}, but is shifted towards $\theta=0$ as the inplane field component $B_{\parallel}$ is turned along the strain direction $y$.
This behavior could be taken as an experimental probe for the merging of Dirac cones.

\section{Concluding remarks}
We have derived the expression of the interlayer magnetoresistance in a multilayer deformed Dirac electron system. We have discussed the signature of the motion and the merging of Dirac cones induced by the deformation.\
 
In the case of independent Dirac valleys, the system shows a negative magnetoresistance for a normal magnetic field as in undeformed case. However, a crossover from a negative to a positive magnetoresistance could take place in the presence of an inplane field component. The latter induces an inplane Lorentz force which reduces the interlayer tunneling. This effect is more pronounced as the amplitude of the deformation is increased.\

The motion of Dirac cones, resulting from the deformation, gives rise to a dependence of the interlayer magnetoresistance on the azimuthal angle. However, the behavior of the interlayer resistance with the out-of-plane angle $\theta$ is unchanged compared to the undeformed case.\

We have argued that the sign change of the magnetoresistance could also result from a coupling between Dirac valleys which removes the degeneracy of the Landau level and, hence, reduces the density of carriers. This coupling is enhanced as the merging is approached or at high magnetic field. A criterion is proposed to define the range of validity of the independent Dirac valleys assumption.\

These features may be observed in $\alpha$-(BEDT)$_2$I$_3$ under high pressure or in a stack of deformed graphene like systems.

\section{Acknowledgment}

We warmly thank G. Montambaux, M. Goerbig, J. N. Fuchs and C. Pasquier for helpful and stimulating discussions. We are grateful to C. Pasquier for providing us with reference \cite{Pasquier} prior to publication.
We are indebted to J.-N. Fuchs and G. Montambaux for a critical reading of the manuscript.
This work was partially supported by the National Research Foundation of Korea (NRF) grant funded by the Korea government (MEST) (No. 2012-0008974) and the Tunisian-French CMCU 10G1306 project.
Parts of this work were carried out in the Max Planck Institut f\"{u}r Physik Komplexer Systeme (MPIPKS) in Dresden (Germany) and CNRS-Ewha International Research Center (CERC) (Seoul, South-Korea).
S. H acknowledges the financial support of (MPIPKS) and the kind hospitality of the members of CERC. 



\begin{references}
\bibitem{Geim} K. S. Novoselov,  A. K. Geim, S. V. Morozov,D. Jiang, Y. Zhang, S.  V. Dubonos, I. V. Grigorieva and, A. A. Firsov, Science, {\bf 306} 666 (2004)
,K. S. Novoselov, A. K. Geim, S. V. Morozov, D. Jiang, M. I. Katsnelson, I. V. Gregorieva, S. V. Dubonos, and A. A. Firsov, Nature {\bf 438}, 197 (2005), 

\bibitem{Review} A. H. Castro Neto, F. Guinea, N. M. R. Peres, K. S. Novoselov and  A. K. Geim, Rev. Mod. Phys. {\bf 81}, 109 (2009)

\bibitem{Kobayashi} A. Kobayashi, S. Katayama, K. Noguchi and Y. Suzumura, J. Phys. Soc. Jpn {\bf 73} 3135 (2004), A. Kobayashi, S. Katayama, Y. Suzumura, and H. Fukuyama,  J. Phys. Soc. Jpn {\bf 76} 034711 (2007).
 
\bibitem{Mark-BEDT} M.O. Goerbig, J.-N. Fuchs, G. Montambaux, and F. Piechon, Phys. Rev. B {\bf 78}, 045415 (2008).

\bibitem{Kino} H. Kino and T. Miyazaki, J. Phys. Soc. Jpn {\bf 75} 034704 (2006).

\bibitem{Pouget} P. Alemany, J. -P. Pouget, and E. Canadell, Phys. Rev. B {\bf 85} 195118 (2012).
\bibitem{Morinari} T. Morinari, T. Himura and T. Tohyama, J. Phys. Soc. Jpn. {\bf 78}, 023704 (2009).
\bibitem{Tajima} N. Tajima, S. Sugawara, R. Kato, Y. Nishio, and K. Kajita, Phys. Rev. Lett. {\bf 102}, 176403 (2009).
\bibitem{Osada08} T. Osada, J. Phys. Soc. Jpn. {\bf 77}, 084711 (2008)
 
\bibitem{Osada11} T. Osada, J. Phys. Soc. Jpn. {\bf 80}, 033708 (2011).

\bibitem{Pasquier} M. Monteverde, M. O. Goerbig, P. Auban-Senzier, F. Navarin, H. Henck , C. R. Pasquier, C. M\'ezi\`ere, and P. Batail, (unpublished).

\bibitem{Kobayashi11} A. Kobayashi, Y. Suzumura, F. Pi\'echon and G. Montambaux, Phys. Rev. B {\bf 84} 075450 (2011).

\bibitem{Piechon} F. Pi\'echon, Y. Suzumura and T. Morinari, cond-mat/1303.2652 (unpublished).

\bibitem{Tarruell}L. Tarruell, D. Greif, T. Uehlinger, G. Jotzu, T. Esslinger, {\it Nature}, {\bf 483}, 302 (2012).

\bibitem{Lim}L.-K. Lim, J.-N. Fuchs and G. Montambaux, Phys. Rev. Lett. {\bf 108}, 175303 (2012).
\bibitem{Bellec}M. Bellec, U. Kuhl, G. Montambaux and F. Mortessagne, Phys. Rev. Lett. {\bf 110}, 033902 (2013).

\bibitem{Gilles09} G. Montambaux, F. Piechon, J.-N. Fuchs, and M. O. Goerbig, Eur. Phys. J. B {\bf 72}, 509 (2009).
\bibitem{Mark} For a review on Dirac cones in deformed graphene see M. O. Goerbig, Rev. Mod. Phys. 83, 1193 (2011).
\bibitem{Himura} T. Himura, T. Morinari, and, T. Tohyama, J. Phys.: Condens. Matter {\bf 23}, 464202 (2011).

\bibitem{Morinari10} T. Morinari and T. Tohyama, J. Phys. Soc. Jpn. {\bf 79}, 044708 (2010).

\bibitem{Esaki} K. Esaki, M. Sato, M. Kohmoto a,d B. I. Halperin, Phys. Rev. B {\bf 80}, 125405 (2009).

\bibitem{math1} F. Marques, O. Negrini and A. J. da Silva, J. Phys. A: Math. Theor. {\bf 45}, 115307 (2012).

\bibitem{math2} F. M. Fern\'andez, arXiv:1204.0229 (unpublished).

\bibitem{Skala} L. Sk\'ala, J. \v{C}\'i\v{z}ek, J. Dvo\v{r}\'ak and V. \v{S}pirko, Phys. Rev. A {\bf 53}, 2009 (1996).

\end{references}
\end{document}